\documentclass[12pt]{iopart} 
\usepackage{graphicx}\begin{document}
\title[Bouncing trimer]{Bouncing trimer: a random self-propelled particle, chaos and periodical motions}
\author{S. Dorbolo, F. Ludewig, and N. Vandewalle}
\address{
   GRASP, Physics Department B5, University of Li\`ege, B-4000 Li\`ege, Belgium}
  

\begin{abstract} The complexity of an object bouncing on a vertically shaken plate resides both in the bouncing process and in the shape of the object.  A trimer is an object composed of three centimetrical stainless steel beads equally distant and is predestined to show richer behaviours than the bouncing ball or the bouncing dimer.  The rigid trimer has been placed on a plate of a electromagnetic shaker and has been vertically vibrated according to a sinusoidal signal.  The horizontal translational and rotational motions of the trimer have been recorded for a range of frequencies between 25 and 100 Hz while the amplitude of the forcing vibration was tuned for obtaining maximal acceleration of the plate up to 10 times the gravity.  Several modes have been detected like e.g. rotational and pure translational motions.   These modes are found at determined accelerations of the plate and do not depend on the frequency.  Chaotic behaviours are observed for other accelerations.   By recording the time delays between two successive contacts when the frequency and the amplitude are fixed, a mapping of the bouncing regime has been constructed and compared to that of the dimer and the bouncing ball.  Period-2 and period-3 orbits have been experimentally observed.  In these modes, according to observations, the contact between the trimer and the plate is persistent between two successive jumps.  This persistence erases the memory of the jump preceding the contact.  A model based on the conditions to obtain persistent contact is proposed and allows to explain the values of the particular accelerations for which period-2 and period-3 modes are observed.  Finally, numerical simulations allow to reproduce the experimental results.  That allows to conclude that the friction between the beads and the plate is the major dissipative process.
\end{abstract}
\submitto{\NJP}
\maketitle

\section{Introduction}
The bouncing ball is a well known problem and is more often than not presented as the example for explaining period doubling route to chaos \cite{TUF,LUC,POL,majumdar,budd,luo}.  Indeed, the bouncing ball problem enlightens how a very basic equation of motion is able to generate chaotic motions.  Recently, the interest in this subject has been renewed through granular shaken systems\cite{martin,swinney}.  When a large assembly of grains is shaken, compaction, convection, or even gas-like behaviours may be observed according to the number of grains and to the acceleration of the plate that vibrates the packing.  The packing may be considered as inelastic bouncing ball \cite{martin}.  However, most of the works concern the description of the bouncing ball or the description of a large amount of bouncing balls.  While the shape of the bouncing particles has been rarely considered.  Actually, the number of possibilities is huge and parameters are numerous.  The difficulty is to fix the relevant parameters.  In the present paper, we have experimentally studied the behaviour of a particular object: a trimer.  That object is made of 3 beads equidistant from each other and linked by rigid rods.  This is not too much far from a ball.  However, this particular shape is relevant in the scope of understanding the bouncing of complex object. 

\subsection{Background}
The contact of a single ball with a plate may be modeled  by one single point of contact despite the complexity of the microscopic bumps present at the surface of both objects in contact.  This simple approach is however very fruitful when the bouncing dynamics of the ball is considered.  The shock between the bead and the plate being characterized by one single parameter: the coefficient of restitution $\varepsilon$ that is given by the ratio between the energy of the ball just before and just after the collision.  This coefficient eventually depends on the impacting speed \cite{IMP1,IMP2}.  When the ball is dropped over the plate, it bounces till inelastic collapse occurs \cite{INE}.   Roughly speaking, when a spherical bead is dropped or excited on a horizontal plate that is vertically shaken according to a sinusoid with an amplitude $A$ and a frequency $f$, it is possible to obtain stable bouncing modes when the speed provided by the plate during the impact compensates the energy loss.   The relevant parameter is the reduced maximum acceleration $\Gamma$ defined by  the ratio between the maximum acceleration of the plate $A \omega^2$ ($\omega$ being the impulsion $A \omega^2$) and the gravity $g$.

The shape of an object is rarely as simple as a ideal sphere.  As a consequence, the object is subject to rotation and the coefficient of restitution depends on the orientation of the object and both its translational and rotational speeds.  As a first step towards a more complex system than a ball, we may think to the ellipsoid.  This object has always one contact with the plate but the contact `surface' depends on the orientation of the object.  Moreover, a new degree of freedom is introduced since rotation is allowed.  However, that is pretty difficult to experimentally study the dynamics of such a well defined shape.  

On the other hand, the increase of complexity may be obtained by increasing the number of contacts between the object and the plate.  The first step, coined dimer, is an object that has two contact points when it is at its stable position on a plate; a dimer is a object formed by two beads linked by a rigid rod.  The bouncing of that object on a vertically shaken plate has been studied in 2004 \cite{PRL,PG}.  The preliminary investigations concerned the bouncing mode when $\Gamma$ is below 1 \cite{PRL}.  In these conditions, the bouncing is obtained only when the dimer is excited by dropping it on the plate.  Different orbits have been observed.  Let us describe them by order of increasing energy.  The mode G (for ground): the dimer does not bounce, both beads remain in contact with the plate.  The mode D (for drift): one of the bead bounces while the other follows the motion of the plate (that latter remains roughly in contact with the plate).  The dimer moves then horizontally.  Its speed depends on the aspect ratio $A_r$ of the dimer, i.e. the ratio between its length and the bead radius.  Namely, the speed decreases with $A_r$.  Compare to the bouncing ball, this mode is totally new since it comes from a additional degree of freedom of the dimer, i.e. the rotation.  The mode J (for jump): both beads bounce together and hit the plate once per period.  The succession of the bouncing of the bead \# 1 and the bead \# 2 follows a complex but periodical sequence.   The mode T (for twist): each bead alternatively hits the plate once every two periods.  

A second paper \cite{PG} investigates what happens at $\Gamma$ above one.  Transitions between modes have been observed.  An interesting description of the bouncing dimer for plate accelerations above $g$ has been developed by Swift {\it et al} \cite{SWI}.  They showed that a bouncing dimer experiences stochastic motion regarding the time delay between successive shocks and the angle at the impact moment.   

The mode D is very interesting because it represents a manner to generate a self-propelled particles without introducing a textured substrate as in \cite{bammert,vongehlen,heinsalu}.  This is particularly useful to study collective behaviours as nicely performed by A. Kudrolli {\it et al} \cite{lumay}.  In this work, spontaneous swirling and clustering are observed.  The analogy between the self-organization of these grains and biological systems is remarkable.  

Along that framework, we are interested to generate self-propelled particles.  Naturally, a trimer composed by three equally distant beads is a good candidate.  Indeed, that co-planar configuration of three beads offers the advantage of being much less sensitive to slight slope of the plate compared to the dimer.  Moreover, we may {\it a priori} imagine that the horizontal motion of the trimer center of mass can be erratic.  The self-propelled trimer could then be considered as a self-propelled random walker.
\subsection{Aims and scope}

The aims of the present work are (i) to determine the forcing oscillation conditions for which the trimer behaves as an erratic self-propelled particle and (ii) to compare the bifurcation diagram of a bouncing ball, a bouncing dimer and a bouncing trimer.   

By way of an introduction, the video file [intro.mov] shows two situations that illustrate both aims of the present work.  Indeed, on the left, a small trimer (see legend) experiences a random motion on the plate.  On the other hand, on the movie on the right, a large trimer bounces randomly before locking a periodic motion (it bounces once every two periods).

In a first step, the inclination of the trimer will be neglected compared to the horizontal motion of the center of mass and the rotations around the vertical axis.  From this simplified point of view, the motion of the trimer can be then decomposed in the rotation and translation motion.  In a second step the vertical motion will be considered.  This later is rather difficult to obtain by image analysis.  That is why we prefer to study the time delay between successive impacts to determine some dynamics of the vertical motion of the beads. 

The experimental set-up is detailed in the following section.  Afterwards, the paper is split into a section concerning the experimental results (Sect. 3) and one about the numerical investigations (Sect. 4).  The Sect. 3.1 is devoted to the center of mass motion and to the rotation of the trimer neglecting the vertical motions.  On the other hand, the Sect. 3.2 is focused around the description of the bouncing mode and the comparison with the bouncing ball and dimer.   Numerical investigations have been performed in order to evidence the relevant parameters governing the stable trajectories and chaotic regimes (Sect. 4).  Finally, the conclusions are drawn.  

\begin{figure} [t]
\begin{center}\label{restitution}
\includegraphics[width=10cm]{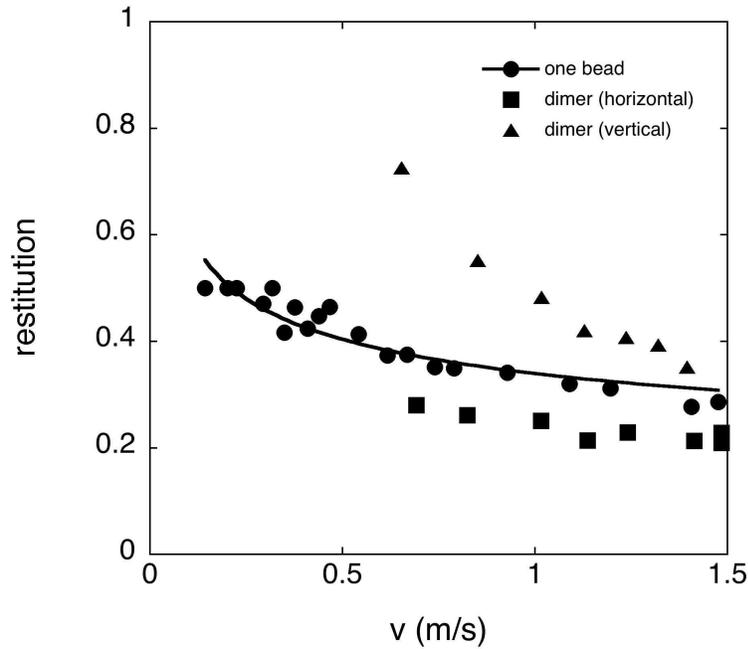}

\end{center}
\caption{Coefficient of restitution of a bouncing ball (circles), a dimer rebounding vertically (triangles) and dimer bouncing horizontally (squares) versus the impact speed of the center of mass. }
\end{figure}

\section{Experimental set-up}
The object under consideration (a bead, a dimer or a trimer) is set on a plate bordered by a squared arena in order to avoid the escape of the object from the plate.  The interaction between the borders and the bouncing object may affect the periodical modes.  Some authors propose to work on a parabolic plate \cite{TUF, anglish}.  Such a system can be envisaged for one single bead but does not hold for dimers and trimers.  For the dimer, a groove should be better than a parabolic plate.  Since no simple solution exists for the trimer, we chose to work on a plane plate and to try to limit the interaction between the bouncing object and the borders.  Both circular and  square arenas have been compared without any noticeable effect on the results.  However, the interaction border-trimer is reduced by glueing a thin disk made of rigid plastic on the top face of the trimer.  The radius of this disk is slightly larger than the trimer.  In so doing, the trimer and the borders have, at worst, only two contact points.

The plate is vertically shaken by an electromagnetic shaker (G\&W V55) via a linear bearing in order to ensure the unidirectionnal vibration.  The considered frequencies are tuned between 25 and 100 Hz while the amplitude allows to reach accelerations up to $\Gamma=10$.  That acceleration is measured thanks to a calibrated accelerometer.  

The motions of the trimer are recorded using a high-speed camera (Redlake).  To begin with, the camera is placed at the vertical position of the plate (results are found in Sect. 5).  That allows to record the horizontal translational and rotational motion of the trimer with respect to the vertical excitation.  A high recording speed is not necessary for this purpose (25 images per second is largely enough).  On the other hand, to analyze the vertical motion, the camera is placed perpendiculary to the axis of vibration and the image acquisition speed is set to 250 images per seconds (results are found in Sect. 6).  

\begin{figure} 
\begin{center} \label{diag}
\includegraphics[width=\columnwidth]{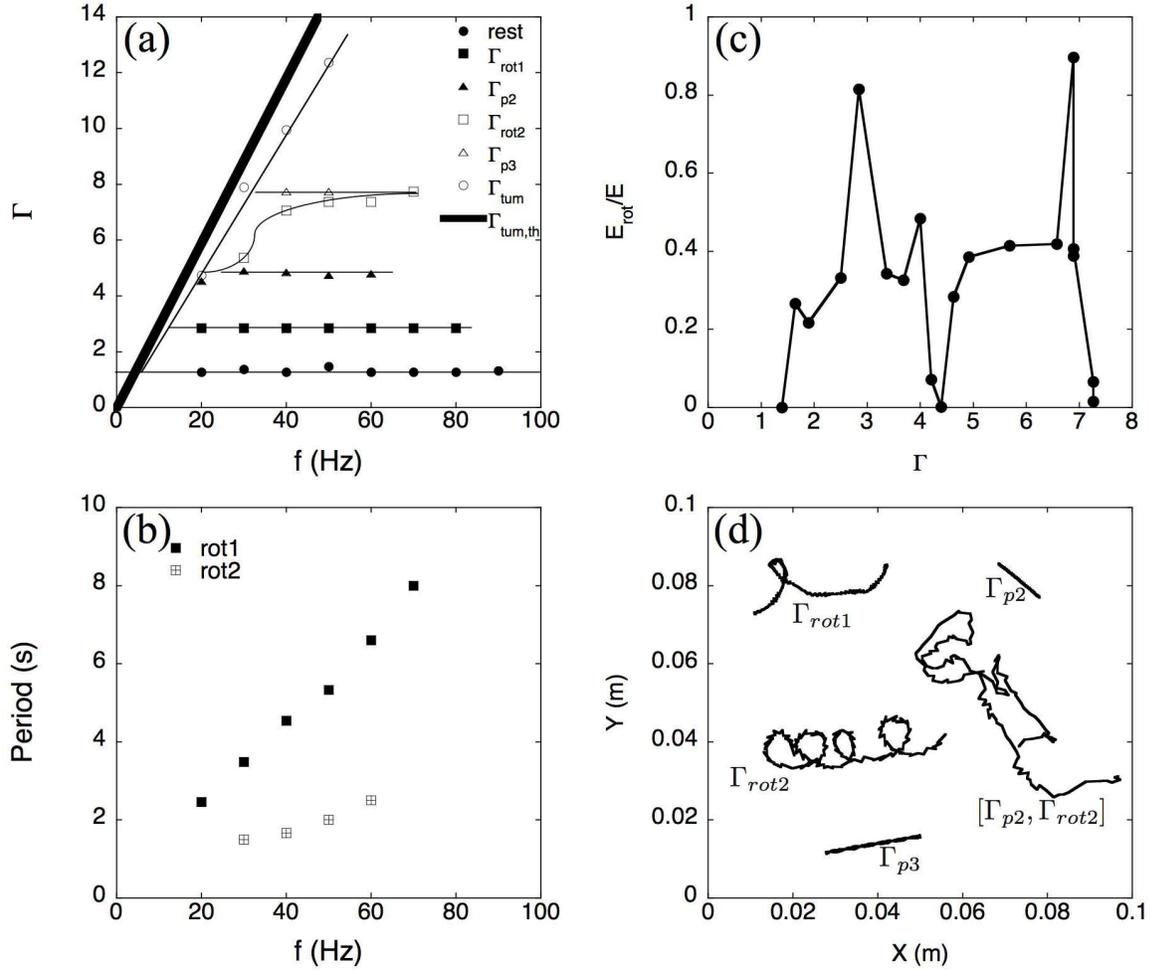}

\end{center}
\caption{(a) Diagram presenting the different modes encountered for a trimer with $A_r=3.5$ when the frequency $f$ and the reduced acceleration $\Gamma$ of the vertical excitation are tuned.  Below the black bullets, the trimer remains on the plate without bouncing.  The trimer locks a periodical orbit along the horizontal lines, namely a rotation mode (black squares), a period-2 mode (black triangles), a second rotation mode (open squares) and a period-3 mode (open triangles).  The trimer tumbles when excited above the oblical line passing by open circles. (b) Comparison of the rotation period in the mode of rotation 1 (black squares) and in the mode of rotation 2 (open squares) as a function of the forcing frequency.  (c) Ratio between the rotational $E_{rot}$ and total energy $E$ of a bouncing trimer at $f=$ 40 Hz. (d) Trajectories of the center of mass of the trimer in the following modes: period-2, rotation-1, rotation-2, period-3 and in a chaos state.  The reduced accelerations indicated on the figure refer to the mode.  See also the movie [intro.mov] and [sweep.mov].}
\end{figure}
Each time that the bouncing object (one bead, a dimer or a trimer) hits the plate, a high and sharp peak is observed on the accelerometer signal.  That allows to measure the time delay $\Delta t$ between two successive shocks.  This method has the inconvenient not to detect simultaneous shocks and not to detect soft shocks (when the bead and the plate collide with a relative speed close to zero).  However, fixing the frequency and the amplitude, a set of time delays ${\Delta t}$ can be accumulated.  The distribution of these time delays allows to extract periodical behaviours as for example a phase locking or a period doubling.  

From the practical point of view, the acceleration of the plate is removed out of the accelerometer signal by using a high-pass filter.  The signal is then recorded by an oscilloscope and transmitted to a computer.  The absolute value of the signal is then cleaned by studying the correlation of the signal and a decreasing exponential $\exp (t/\tau)$ where $\tau(\approx 1$ ms$)$ represents the typical relaxation of the signal after a shock, $t$ is the time.  A peak is considered as resulting from a shock when the peak value is larger than a preliminary determined threshold $P$ that depends on $f$ and $A$.  The function $P(f,A)$ is basically determined by studying the accelerometer signal without any bouncing object on the plate.

To sum up, the measurement procedure is the following.  The frequency is fixed to $f$.  The amplitude range of the plate is chosen so as to sweep reduced accelerations between 0.5 and 10 by a certain amplitude step $\Delta A$.  To begin with the signal from the accelerometer is recorded and treated when no object is present on the plate.  That allows to determine the threshold values $P (A,f)$.  After that first run, the shaker is stopped and the object placed on the plate.  Data sets of time delays ${\Delta t}(A,f)$ are collected for each value of the tuned amplitude.  From these sets, the probability distributions of the $\Delta t$'s are deduced.  The density of probability of $\Delta t$ is mapped on a ($ \Delta t, \Gamma$) plot (the frequency being fixed).

\section{Experimental results}
\subsection{Coefficient of restitution and choice of sizes}
For a bead, the characterization of the bouncing is mainly governed by the coefficient of restitution $\varepsilon$.  That coefficient is given by the ratio between the speed of the object just before the impact and the speed just after the impact.  That coefficient is quite easy to measure for one single bead.  For a dimer, things turn to be more complex since the orientation of the dimer before the impact and the rotations have to be considered.  It is pretty difficult to define this coefficient for the trimer. Indeed, when the object is more complex, the object tangential speed with the plate may be non zero.  The main consequence is that friction between the bead and the plate must be taken into account.  The energy loss processes are the elastic restitution and the friction.  

As first measurements,  the restitution coefficient for one single bead and for a dimer ($A_r=3.5$) have been studied.  The dimer has been released horizontally and vertically on the plate in order to evidence the range of restitution coefficients.  Using an electromagnet and a high speed camera located far enough from the impact place, it is quite convenient to determine the speed of the center of mass before and after the impact.  The use of the electromagnet allows to diminish rotation of the dimer before and even after the collision since the horizontality (or verticality) is conserved before and after the impact.  In Fig.1, the coefficient of restitutions are represented.  The circles, squares and triangles are for the single bead, the dimer that impacts the plate horizontally and the dimer that impacts the plate vertically respectively.  

The coefficient of restitution of the single bead is found to increase when the impact speed is decreased.  Our measurements are consistent with theoretical behaviour of $\varepsilon(v)$ found in ref.\cite{IMP1} and are scaled by $v^{-1/4}$ (continuous line in Fig.1).  For the single bead, $\varepsilon$ varies between 0.4 when the impact speed is 0.1 m/s and 0.3 when the speed is larger than 1 m/s.  

It is noticeable that the dimer when hitting the plate horizontally has the lowest coefficient of restitution of the three considered cases.  Indeed, that can be understood as considering the double contacts of the dimer with the plate.  At the precise moment of this double events, both beads generate a rotational momentum to the dimer.  The momenta have a different sign that has for consequence to frustrate the dimer, in other words a lot of energy is dissipated in the elastic deformation of the dimer.  On the other hand, when the dimer is vertically released, the coefficient is about twice the coefficient of the horizontal case and thus larger than the single bead.  The large restitution observed with the vertical dimer is attributed to the elasticity of the rod.

The experiments realized with bouncing dimers learn us that the speed (the motility) of such a particle depends on the aspect ratio of the dimer \cite{PRL}.  According to these measurements, the maximal horizontal drift speed of the dimer is found when $A_r=3.5$.  The speed decreases to zero at $A_r=5.5$.  In Table I, the speed of the dimer versus the frequency are given.  

In order to evidence the aspect ratio role, two trimers have been constructed with aspect ratios equal to 3.5 and 5.6 respectively.  The centimetrical stainless steel beads are glued on a polycarbonate triangle which mass may be neglected compared to the mass of the beads (4.1 g per bead).
The mass $m$ of the trimers are 14.6 g and 18.5 g respectively for the trimer with $A_r=3.5$ and 5.6.  The mass of the plexiglas supporting plates are then 2.3 g and 6.2 g respectively. 
\begin{table}\begin{center}
\begin{tabular}{| c | c | c |}
\hline
$A_r$ & frequency & speed (cm/s)\\ \hline 
3.5  & 25 & 2 \\
 \/ & 50 & 1 \\
 \/  & 75 & 0.7 \\Ê
 5.5  & from 25 to 75 & $<0.1$\\ \hline
\end{tabular}
\caption{Horizontal drift speed of a dimer.  The reduced acceleration is 0.9, the aspect ratio is indicated.}
\end{center}
\end{table}

\subsection{Tracking the trimer motions}

To begin with, we vertically shake a trimer with $A_r=$3.5 because this size corresponds to a very mobile dimer aspect ratio.  The diagram of Fig.2a represents the different encountered modes with respect to the frequency $f$ and the reduced acceleration $\Gamma$.  To enumerate the different modes, the behaviour of a trimer shaken at 40 Hz will be described here after.   Starting from $\Gamma=0$, the trimer remains in contact with the plate until $\Gamma_{rest} \approx 1.3$.  The trimer bounces randomly till $\Gamma_{rot1} \approx 3$.  At that acceleration, it rotates on itself, the horizontal translational energy is zero.  Increasing slightly the acceleration conducts the trimer to a random translational and rotational motion.  When $\Gamma_{p2} \approx 5$ is reached, the trimer bounces once every two periods.  Seen from the top, the horizontal translational and rotational motions are stopped.  The plane of the trimer remains parallel to the plate.  Increasing again the acceleration, the random motion is again observed till $\Gamma_{rot2} \approx 7$.  There, a rotation mode is observed which rotation is larger than the previous rotation mode at $\Gamma=\Gamma_{rot1}$.  However, this mode is not very stable and difficult to obtain.  At $\Gamma_{p3} \approx 8$, the trimer bounces once every 3 periods, the plane defined by the three beads remains parallel to the plate.  Finally, the trimer tumbles for $\Gamma_{tum} > 10$.  The limit acceleration $\Gamma_{rest}$, $\Gamma_{rot1}$, $\Gamma_{p2}$, $\Gamma_{rot2}$, and $\Gamma_{p3}$ are represented on Fig.2a. It is remarkable that these particular accelerations do not depend on the frequency.  That experiment also shows the windows of stability of the bouncing trimer system that does only depend on the reduced acceleration $\Gamma$.   On the other hand, the acceleration for tumbling is observed to vary linearly with the frequency. 

A simple argument allows to explain that.  The trimer may tumble when one of the bead can get enough kinetic energy from the plate to reach a height of at least the length of the bissector of the triangle defined by the beads of the trimer.  The maximum accessible speed for a bead can be taken as the maximum speed of the plate $A \omega$.  The balance between the kinetic energy and the potential energy writes
\begin{equation}
\frac{1}{2}m (A_{tum} \omega)^2=m g h
\end{equation} where $h$ is the length of the bissector $h=\frac{\sqrt{3}}{2} (A_r-1) r$ and  $A_{tum}$ is the plate amplitude for which the trimer tumbles.  Injecting the expression of the amplitude with respect to the reduced acceleration, we obtain
\begin{equation}
\Gamma_{tum,th}= \omega \sqrt{\frac{h}{g}}
\end{equation} $\Gamma_{tum,th}$ being the theoretical reduced acceleration for tumbling.  That law is represented as a thick line on the Fig.2a and fits pretty well the experimental data.

The file [sweep.mov] shows the dynamics of the trimer when the acceleration is slowly increased from zero to about 5 $g$.  The rotation and period-2 modes are clearly visible (the frequency was fixed at 25 Hz).

The rotational mode is a kind of generalization of the drift motion observed for the dimer.  The propulsion is a consequence of the friction between the beads and the plate \cite{PRL}.  The periods of rotation have been measured for the rotation-1 and -2 that occur respectively at $\Gamma \approx 3$ and 7.  They are presented in Fig.2b versus the excitation frequency.  A period of about 2 s corresponds to a linear speed of a bead of about 3 cm/s.  That is compatible with the measurement of the dimer.  The plain and open squares are for the mode 1 and mode 2.  The period of rotation linearly increases with the frequency.  Indeed, that slowing has been also observed with the dimer.  Indeed, the linear speed of a dimer scales with the inverse of the frequency as demonstrated in \cite{PRL}.

 In order to evaluate the energy $E$ of motion (neglecting the vertical contribution), the translational $E_{tran}$ and the rotational $E_{rot}$ energy have to be calculated:
\begin{equation}
E=E_{tran}+E_{rot}=\frac{1}{2}M v^2+\frac{1}{2}I \dot \phi^{2}
\end{equation} where $M$ is the total mass of the trimer, $I$ is the angular momentum and $\dot \phi$ is the rotation speed of the trimer.  The angular momentum is given by :
\begin{equation}
I=3 (m d^2+\frac{2}{5}m r^2)
\end{equation} where $m$ is the mass of one bead, $d$ the distance between the center of mass of the trimer and the center of mass of one bead and $r$ the radius of a bead.  The energy (horizontal) has been computed thanks to the recording of the trimer trajectories.  In Fig. 2c, the ratio between the rotational kinetic energy $E_{rot}$ and the total kinetic energy $E$ is plotted with respect to the reduced acceleration $\Gamma$ while the frequency is kept constant at 25 Hz.  The peaks at $\Gamma=3$ and 7 corresponds to the rotation modes while the zero found at $\Gamma=4.5$ and 7.3 correspond to period-2 and period-3 modes respectively.  It is noticeable that the most chaotic motion is found between $\Gamma=5$ and 6.5.  In that range of forcing acceleration, the rotational energy is nearly equal to the translational one.  Self-propelled random walker particles are thus found in that range of acceleration.

Finally, the trajectories have been recorded during 40 s for 5 particular accelerations: $\Gamma_{rot1}<\Gamma_{p2}<\Gamma_{walk}<\Gamma_{rot2}<\Gamma_{p3}$ where $\Gamma_{walk}$ is a reduced acceleration of the plate intermediate between $\Gamma_{p2}$ and $\Gamma_{rot2}$.  The trajectories are represented in Fig.2d, $X$ and $Y$ being the coordinate of the center of mass of the trimer on the plate.  The period-2 and period-3 modes are characterized by a linear trajectory.  The speed is very low (0.5 mm/s).  On the other hand, the trajectories for both rotational modes are loops.  The horizontal speeds are again small ($\approx$ 1 mm/s).  The trimer when accelerated at an acceleration between $\Gamma_{p2}$ and $\Gamma_{rot2}$ has a horizontal speed of 5 mm/s.  During the period of time of the measurement (40 s), the trimer randomly turns on itself.  That shows that the self-propelled random walker is found in that range of reduced acceleration.  However, the recording time is too small to allow to determine whether the motion is under or over-diffusive.

\subsection{Mapping the trimer bouncing modes}
Some modes may be not observed by the eye because they have long period or are very fast.  The time delays between two successive shocks of any beads on the plate have been chosen to map the system with respect to the reduced acceleration while the frequency is fixed.  The procedure used to obtain these diagrams is described in the Sect.2.  The frequency has been fixed to 25 Hz since the period is rather large, that allows to have a maximum of time resolution.  In Fig.3, the bouncing ball (top), the bouncing dimer ($A_r=3.5$) (middle) and the bouncing trimer ($A_r=5.6$) (bottom) are compared.  The darker is the plot, the more probable is the time delay between successive shocks.  The experimental method is quite similar to that used in the oscillation of a thin layer of grains \cite{swinney}.
\begin{figure} [t]
\begin{center}
\includegraphics[width=10cm]{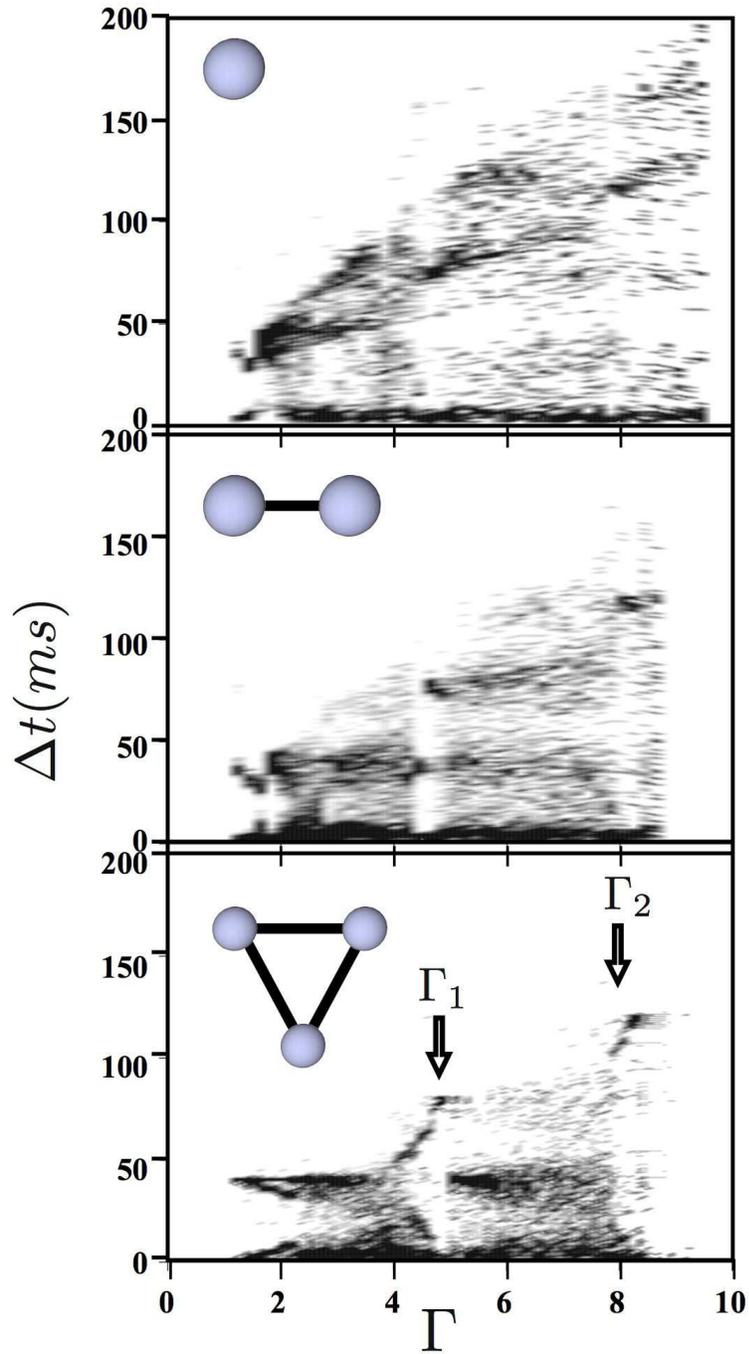}

\end{center}
\caption{Density of probability of finding a time delay between two successive bead-plate impact with respect to the reduced acceleration.  The darkest is, the most probable.  The frequency is fixed at 25 Hz (period=40 ms).  Three cases are compared: the bouncing ball (top), the dimer ($A_r=3.5$) and the trimer ($A_r=5.6$).  The arrows indicate the particular values $\Gamma_1$ and $\Gamma_2$.}
\end{figure}
\begin{figure}[t]
\begin{center}
\includegraphics[width=10cm]{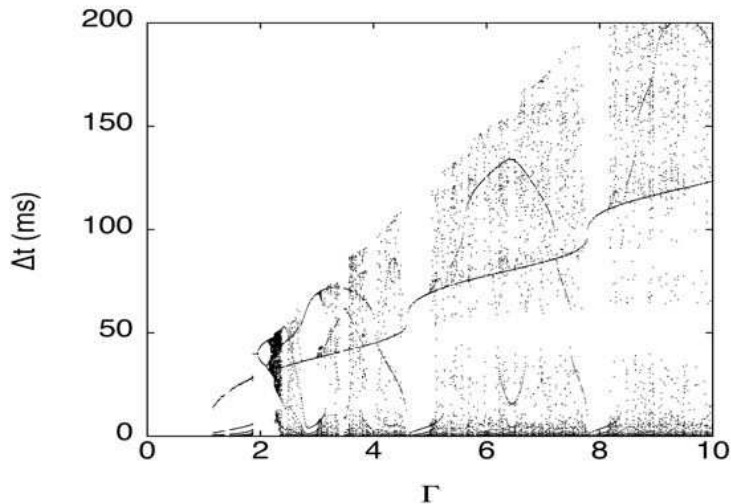}

\end{center}
\caption{Theoretical map of the time delay between successive impacts for a bouncing ball with a coefficient of restitution of 0.35.}
\end{figure}

Let us begin by describing the bouncing ball diagram.  Since only one single ball is concerned, the time duration between two successive shocks with the plate represents also the flight time of the bead.  This flight time is a well known parameter and is theoretically related to the logistic map \cite{LUC}.  More often than not, the completely inelastic bouncing ball diagram is represented because it evidences very clearly the period doubling route to the chaos.  In our case, the bouncing is characterized by a certain coefficient of restitution $\varepsilon$, defined as the ratio of the ball speed just before and just after the collision with the plate at rest.  That coefficient has been measured with respect to the impact speed.  Indeed, we observe an increase of the coefficient when the impact speed goes to zero \cite{IMP1,IMP2}.  The mean value for $\varepsilon$ has been found to be 0.35.  It is very easy to compute the flight time of the bouncing ball according to the reduced acceleration of the plate when the frequency is fixed.  Fig.4 presents the results from the simulation for the bouncing ball with $\varepsilon=0.35$ and $f=25$ Hz.  The general shape is in good agreement with the experimental bifurcation diagram, especially according to the stability windows and according to the forbidden flight times.  On the other hand, the period doubling cascade cannot be experimentally observed because the time resolution is not small enough.

\begin{figure} [t]
\begin{center}
\includegraphics[width=\columnwidth]{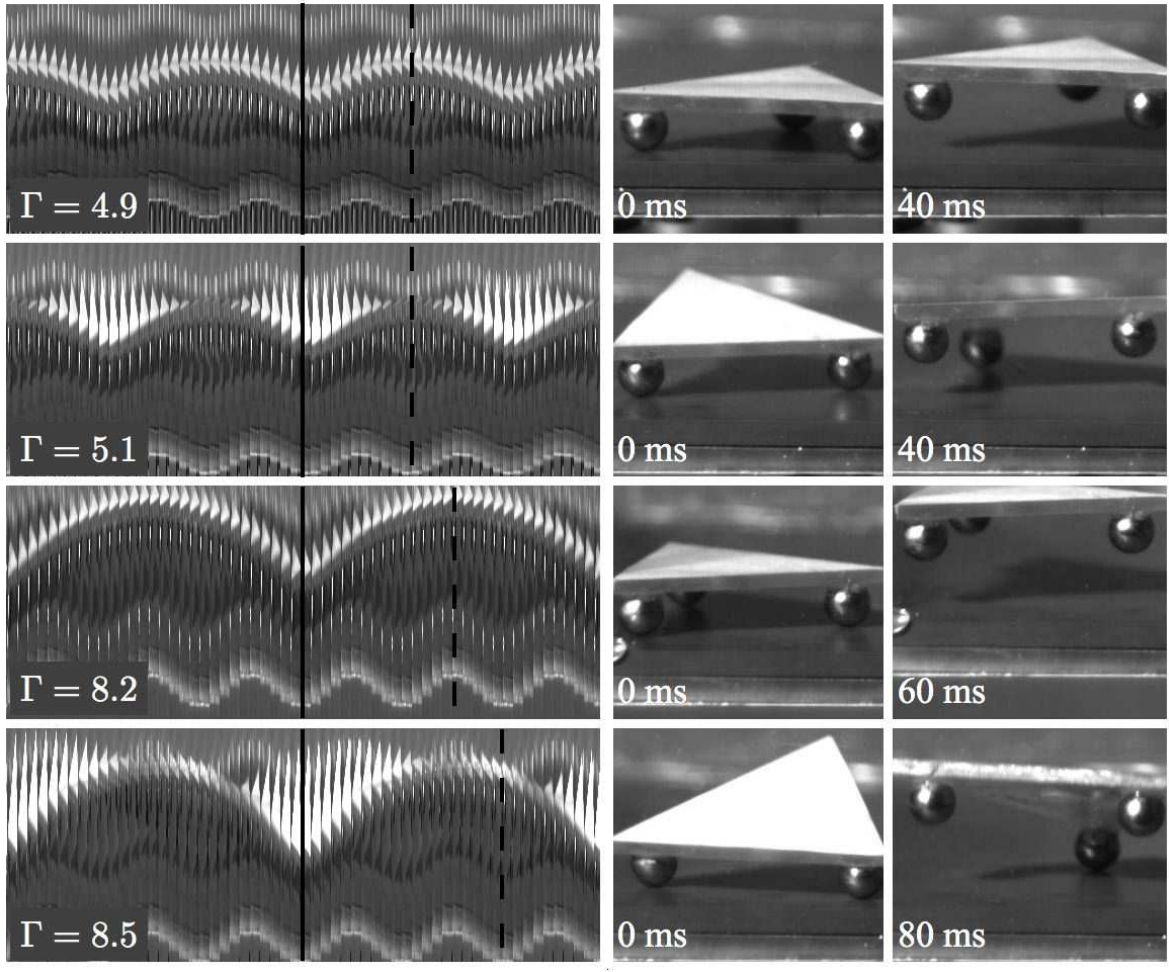}
\caption{The figure aim is to compared the trajectories of the trimer and the plate.  From top to bottom, the trimer is in the period-2 mode, in the twist-2 mode, in the period-3 mode and finally in the twist-3 mode.  The reduced accelerations are indicated on the left. The 2 pictures on the right are snapshot of the trimer (times are indicated). For each case, a series of 60 snapshots have been juxtaposed, the snapshots having been horizontally shrunken.  The sinusoid curve at the bottom of each picture in the left column represents the motion of the plate while the white sinusoidal-like zone represents the motion of the trimer.  The more white, the more a top view of the trimer is seen.  In other words, the white colour indicates the inclination of the trimer.  The vertical plain and interrupt lines indicates the place of the snapshots from the central column and from the left column respectively.  See also the movie [diag.mov] that illustrates the different observed periodical motions.}
\end{center}

\end{figure}
When comparing the ball, the dimer and the trimer diagrams, we observe that generally the time delays between shocks are decreased.  That first observation is only natural since the number of beads is increased.  Secondly, two stability windows are encountered at $\Gamma \approx 5$ and $\Gamma \approx 8$.  At these accelerations, period-2 and period-3 orbits are observed.  They are characterized by a darker area at ($\Gamma=5,\Delta t=80$ ms) and ($\Gamma=8,\Delta t=120$ ms).  Note that (i) at these accelerations, the dimer twists from one bead to the other and (ii)  the rotational modes are not observed with that large trimer.

The trimer diagram is the most rich according to the structures that can be observed.  No time delay larger than 40 ms is observed for accelerations till $\Gamma =4$.  It is interesting to observe that below $\Gamma=2$, three branches exist: one is constant at $\Delta t_{branch \ 1}$=40 ms, the two others are complementary: $\Delta t_{branch \ 2}+\Delta t_{branch \ 3}=$40 ms.  The bouncing order of the three beads is difficult to determine even at $f$=25 Hz.  The dynamics of the contact is pretty delicate to experiment.  The motion of the trimer makes an automatic optical detection difficult mainly because the spatial size of the contacts is much lower than the size of the trimer.

Actually, two additional modes have been detected.  They occur for accelerations just above $\Gamma_{p2}$ and $\Gamma_{p3}$.  The trimer bounces alternatively on two of the three beads and then on the third one.  It results a motion of twist of the entire trimer.  Moreover, an horizontal and rectilinear motion is then observed.  The signature of one of these twist modes is seen as darker area for acceleration just above $\Gamma \approx 5$.  

The period-3 orbit is interesting because 'period-3 is a clue for chaotic motion' \cite{li-york,sarkovskii}.  The motion of the center of mass for reduced accelerations between $\Gamma_{p2}$ and $\Gamma_{rot2}$ for the small trimer is certainly chaotic and the random walker characteristic is well checked.

A more precise description of the system is shown in Fig.5.  Pictures of four particular modes have been taken using the high speed camera.  From the top to bottom, the period-2 mode, the twist-2 mode, the period-2 mode and the twist-3 mode are represented (the acceleration are indicated on the left).  The two pictures on the right are snapshots taken during the bouncing.  The time are indicated on the Figures.  The left side of Fig.5 are pictures composed by the juxtaposition of 60 successive snapshots as the ones shown on the right.  They are separated by 4 ms.  The snapshots have been shrunken along the horizontal axis in order to evidence the trajectory of the trimer and the plate.  The plate motion is the sinusoidal curve at the bottom of each picture.  The trimer motion can be seen as the white sinusoids.  The solid and the dashed vertical lines represent the snapshot from the second column ($t=$ 0 ms) and from the third column respectively. 

The period-2 and period-3 modes are clearly visible along the first line and the third line of the Fig.5.  It is remarkable to observe that the trimer experience a jump of more than 2 centimeters before the three beads  hit the plate simultaneously, the trimer plane being parallel to the plate during the whole flight.  The contact between the plate and the beads is a bit particular for both modes.  That is particularly clear when listening to the noise produced by the shocks when the trimer hits the plate.  Indeed, when the period-2 or period-3 modes are obtained, the noise decreases drastically.   The trimer comes in contact very softly and looks more like a delicate landing.   That means that the trimer and the plate have approximately the same speed. This can be due to the inelastic collapse of the trimer on the plate \cite{LUC,majumdar}, even if we do not have direct proof of chattering.  The contact lasts also more than 10 ms.  According to that observation, the rebound looks like completely inelastic from the point of view of the plate.  The memory of the previous bounce is completely erased.

The modes called twist-2 and twist-3 are the ones characterized by a succession of a two beads contact followed by the third bead contact.  In Fig.5, the twist is observed through the variation of the white area according to the inclination of the trimer.  In the twist-3 mode, the time delay between the double impact and the single one is twice of the time delay between the single impact and the double.  This asymetry is not observed with the twist-2 mode but that may be due to a lack of time resolution.  The movie [diag.mov] shows the modes period-2, twist-2, period-3, and twist-3 and their positions in the phase diagram $(\Gamma,\Delta t)$.

The high speed camera recordings have shown that the contacts between the trimer and the plate are persistent in the period-2 and period-3 modes.  Moreover, the intensity of the hits is decreased in these modes, the noise generated by the collision being lowered in these modes.  The persistence of the contact is the ingredient that explains why the modes are observed at a given reduced acceleration that does not depend on the frequency (see Fig.2a).  Indeed, when the contact persists, the memory of the former jump is erased.   The conditions for obtaining a persistent contact is that the trimer and the plate must have the same speed when the trimer comes in contact with the plate.  When this condition is fulfilled, the trimer does not bounce and remains in contact with the plate until the acceleration of the plate becomes smaller than $-g$.  The equations of motion of the plate are \begin{eqnarray}
z & = & A \sin (\omega t) \\
\dot{z} & = & A \omega \cos (\omega t) \\
\ddot{z} & = & - A \omega^2 \sin (\omega t) 
\end{eqnarray} where $z$ is the vertical coordinate.  The take-off phase $\delta$ is deduced from
\begin{equation}
-A \omega^2 \sin (\delta)=-g
\end{equation} The take-off phase is then equal to $\delta=\arcsin (\Gamma^{-1})$.  The trimer and the plate have at the take-off moment the same same vertical position $z_0$ and the same speed $v_0$ 
\begin{eqnarray}
x_0 & = & \frac{g}{\omega^2} \\
v_0 & = & \frac{g}{\omega}\sqrt{\Gamma^2-1}
\end{eqnarray}  It is convenient to chose $t=0$ at the take-off phase.  The equation of motion of the plate becomes 
\begin{eqnarray}
z & = & A \sin (\omega t + \delta) \\
\dot{z} & = & A \omega \cos (\omega t +\delta) 
\end{eqnarray} After the take-off, the trimer experiences a parabolic flight until it hits the plate again.  We impose that the speeds of the trimer and of the plate are same at the moment of this contact.  These constraints are written as following
\begin{equation}
\begin{array}{ccccccccc}
1 & + & \sqrt{\Gamma^2-1}\ & \omega t & - &\frac{1}{2} & (\omega t)^2 & = & \Gamma \sin (\omega t + \delta) \\
\sqrt{\Gamma^2-1} & - & & \omega t  &  & & & = & \Gamma \cos (\omega t +\delta)
\end{array} \end{equation} The Eq.(13) expresses the collision between the trimer and the plate and the Eq.(14) imposes that the speed of the trimer is the same as the speed of the plate.  By taking the squares of both equations and after addition, the solution for $\omega t$ is very simple
\begin{equation}
\omega t = 2 \sqrt{\Gamma^2-1}
\end{equation}  Injecting Eq.(14) in Eq.(13), we find
\begin{equation}
\sin (2 \sqrt{\Gamma^2-1} + \arcsin (\Gamma^{-1}))=\Gamma^{-1}
\end{equation} that is only possible if $2 \sqrt{\Gamma^2-1} = (2 n +1)\pi$ where $n$ is an integer different from 0.  In other words, the persistent contact modes that are periodical orbits occur only for specific reduced accelerations $\Gamma_{n}$ given by $$\Gamma_{n}=\sqrt{\frac{(2 n+1)^2}{4}Ê\pi^2+1}$$  This formula is valid for ball, dimer and trimer.  The period-1 (not observed) occurs at $\Gamma_0=1.86$ while period-2 and period-3 should occur at $\Gamma_1=$4.82 and $\Gamma_2=$7.92 respectively.  These values are indicated by arrows in Fig.3 showing the good matching with the experimental data.  Before all, this model explains why period-2 and period-3 modes does occur at a certain reduced acceleration.  Neither the frequency nor the coefficient of restitution intervene in the determination of the accelerations requested for obtaining these periodical modes.

\section{Numerical simulation}
\subsection{Numerical method}
The numerical simulations helps in the determination of the phases at which a bead of the trimer hits the plate.   In order to obtain the contact phase map, the main difficulty that simulations have to tackle is the multiple contact management.  Especially, the interaction between the resulting forces after the collision may generate frustration that is pretty hard to resolve numerically.   The determination of the restitution coefficient illustrates this problem.  When a single ball is concerned, it is very easy to define the restitution: the speed before and after the shock can be measured and the ratio determines the coefficient of restitution that eventually depends on the initial speed.  On the other hand, the situation is more complicated when a multiple bead object is considered.  For a dimer, $\varepsilon$ depends on the angle that the dimer makes with the horizontal at the moment of the shock.  The coefficient of restitution depends on the angle, the angular speed and the speed of the center of mass.  For the trimer, another degree of freedom has to be added.  The energy dissipation due to the shock between a multiple bead object and the substrate finds it origin in the elastic dissipation when the trimer is bent and the friction between the beads and the substrate.  

\begin{figure}[htbp]\begin{center}
\includegraphics[scale = 0.3]{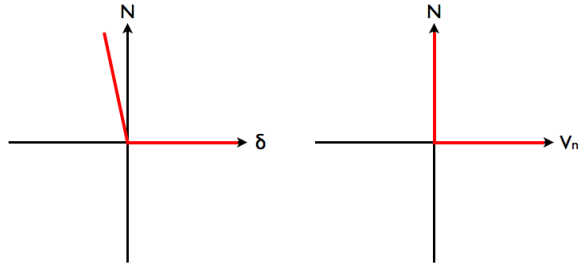}
\caption{Illustration of Signorini's diagrams. The left diagram is smooth and describes a relation between normal force $N$ and distance between the body $\delta$. The right diagram is non-smooth and presents a relationship with normal relative velocity $v_n$ instead of gap $\delta$.}
\label{signorini}\end{center}
\end{figure}

Instead of other models of Discrete Element Method (DEM), the Non-Smooth Contact Dynamics (NSCD) solves the contact forces in the local bases of the contacts \cite{ludwig}. The force is considered as the sum of the normal $N$ and the tangential $T$. The normal force is usually estimated via the smooth Signorini diagram illustrated on the Fig.\ref{signorini}. The normal force is evaluated as a function of the gap $\delta$ (distance between the bodies in contact).

The particularity of the NSCD is the use of a Signorini diagram which links normal force to the relative normal velocity $v_n$ between the body in contact. Some hypothesis must be verified for this purpose. Moreover, the Signorini diagram is non-smooth. The main consequence of that is the existence of a wide indetermination when the relative velocity is zero. To avoid this indetermination, another relation between normal force and relative velocity must be built (see below).
The tangential force is estimated via the Coulomb diagram. In this case, the model deals with velocity.  The indetermination appears when the tangential relative velocity is zero, i.e. when the body rolls without sliding or in the static case. For the usual Coulomb diagram, the tangential force is always null when the relative velocity is null as illustrated in the Fig.7.

\begin{figure}[htbp]\begin{center}
\includegraphics[scale = 0.3]{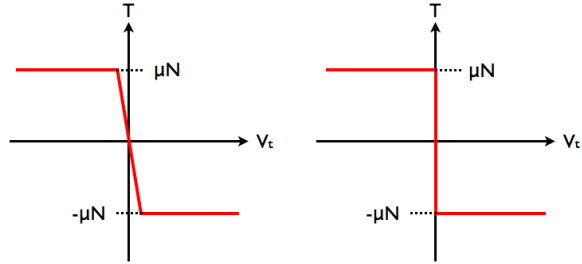}
\caption{Illustration of Coulomb's diagrams. This diagram shows the relationship between the tangential force to the tangential relative velocity. The left diagram is smooth and the right one is non-smooth representation.}
\label{coulomb}\end{center}
\end{figure}

To avoid the indetermination on both normal and tangential components of the contact force, another relationship between the component and, the relative velocity must be established. These relationships are based on the Newton's equations.  The trimer is simulated by obliging the sphere to remain at a constant distance from each other.  In so doing, there is no need of an explicit description of the inertia.  The simulation procedure is the following.  The frequency and the amplitude of the plate are first fixed.  The trimer is released with a slight angle with the horizontal plane in order to break the symmetry.  The simulation runs with a time step of 4 $\mu$s .  After 5 seconds of simulation, the time delays between shocks of the bead with the plate are recorded.  Moreover, it is possible to distinguish which bead hits the plate.  Finally, the phase is also recorded in order to better understand the motion of the bouncing trimer.

\subsection{Results of the simulations}

\begin{figure} [t]
\begin{center}
\includegraphics[width=13cm]{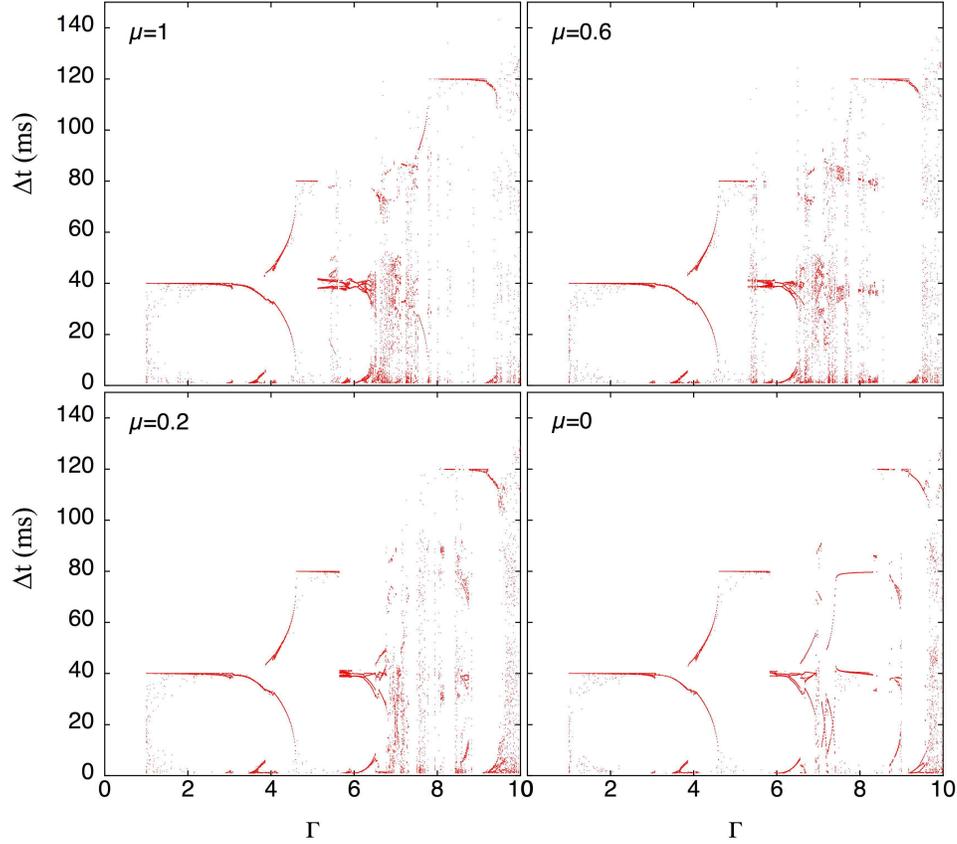}
\caption{Comparison between numerical simulation results for four different friction coefficient: 0, 0.2, 0.6 and 1 while the coefficient of restitution of the bead is zero. The forcing frequency is 25 Hz. }
\end{center}

\end{figure}

 An example of the numerical bouncing dimer is shown in the movie [num.mov] for the following set of parameters: $A_r=5.6,\  f=25$ Hz $,\  \Gamma=9.69, \varepsilon= 0,\  \mu=1$.  The movie on the left is recorded from the front while the movie on the right is the view from the top in order to emphasize the horizontal excursion of the trimer.

The trimer with an aspect ratio of 5.6 has been tested for different values of the coefficient of restitution and of friction.   To begin with, let us describe the result for $\varepsilon=0$ and $\mu=1$ represented in Fig. 8a.  The set of the $\Delta t$ between successive shocks of any bead of the trimer is plotted versus the reduced acceleration.  The diagram is very similar to the experimental results (Fig.3 bottom): (i) a plateau at 40 ms (25 Hz) for low acceleration, (ii) a bifurcation at $\Gamma=4$, (iii) a period-2 plateau at $\Gamma=5$, (iv) a large concentration of 40 ms events at $\Gamma=6$, (v) a chaotic zone between $\Gamma=7$ and 8, (vi) and finally a period-3 plateau at $\Gamma=8$.

\begin{figure} [h]
\begin{center}
\includegraphics[width=13cm]{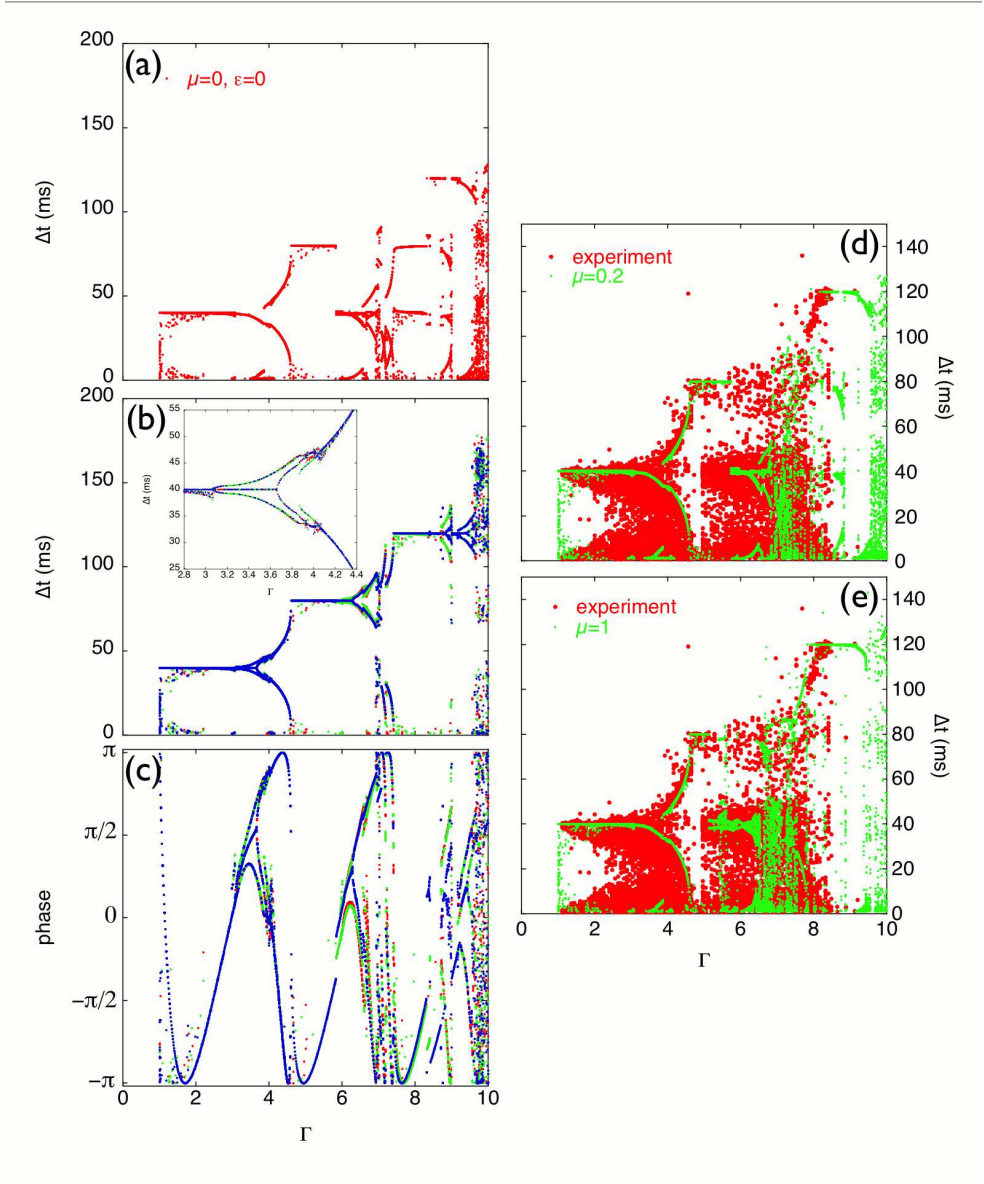}
\caption{Comparison between the experimental and the simulation results.  (a), (b), (c) Simulation of a bouncing trimer which aspect ratio is 5.6, the coefficient of restitution of one bead is zero while the coefficient of friction is zero.  The forcing frequency is 25 Hz.  The Fig. 9a is the classical $(\Gamma,\Delta t)$ (Fig.8d).  In regards of this Figure, the sets of the time delay between two shocks for each particular bead is represented in Figure b.  The different colours corresponds to the different beads.  Finally, the Figure c represents the phase of the plate at which each bead bounces.  Comparison of the experimental (top) and the simulation (middle) sets of time delay $\Delta t$ with respect to the reduced acceleration $\Gamma$.  The Figure d and e compare the numerical simulations $(\varepsilon=0, \mu=1)$ and  $(\varepsilon=0, \mu=0.2)$ to the experimental results (all the points are shown. }
\end{center}

\end{figure}

The modification of the coefficient of restitution has shown that this coefficient does not plays a role.  It does not change drastically the shape of the diagram $(\Gamma, \Delta t)$.  On the other hand, the coefficient of friction does.  In Fig. 8, the $(\Gamma, \Delta t)$ diagrams are represented for 4 friction coefficients: $\mu=1$, 0.6, 0.2, and 0.  The diagrams are the most complex when $\mu$ is high.  Indeed, the dissipation is the highest and the rolling motion is predominant.  The stability windows have slightly moved and are larger for low values of $\mu$.  Increasing the friction moves the stability windows towards lower reduced acceleration.  The interpretation of this study reveals that the restitution and the dissipation laws of such a complex object have to be carefully determined.  However, within this model, the general behaviour of the trimer has been very finely reproduced.

In Fig.9a, b and c, the bouncing of the trimer is analyzed from (a) the point of view of the sets of time delay between two successive contacts of any bead with the plate (as in the experimental process), (b) the time delays between two successive contacts with the plate for each individual bead (on colour per bead) and (c) and the phases when a bead hits the plate (one colour per bead).  The parameters used for the simulation are the frequency at 25 Hz, a coefficient of restitution (of one bead) equal to 0.  The comparison between the behaviour of the trimer and the individual beads allows to show that there exist degenerate modes.  For example, around $\Gamma=8$, all the beads hit the plate once every 3 periods but at different phases.  It results that the time delay between two successive shocks may exhibit only one branch (above $\Gamma=8$) or 3 branches (below $\Gamma=8$).  The study of the individual bounces shows period doublings and bifurcations (emphasized by the insets).  The beads does not play the same role and the initial conditions are of importance.  

A direct comparison has been attempted using the experimental measurements for the restitution and the friction, $\varepsilon=0.35$ and $\mu=0.2$.  In the Fig. 9d and 9e, the red points represent the time delay measurements for the trimer with respect to the acceleration $\Gamma$.  It is from these measurements that the density plot (Fig.3 bottom) has been deduced.  Numerical simulations have been performed with the same restitution and friction and the collection of the time delays $\Delta t$ are reported in Fig.9c.  The behaviours of both sets of time delays are quite consistent.  However, a better agreement between experiment and theory is obtained when the chosen parameters are $\varepsilon=0.35$ and $\mu=1$.  That means that the dissipative processes due to the elasticity of the junctions between the beads should have been considered.

Finally, as shown in the movie [num.mov], the numerical simulations are able to reproduce the random horizontal motions of the trimer.  In Fig. 10, the trajectory of a 3.5 aspect ratio trimer is represented when it is shaken at $\Gamma=6.6$ and at a frequency of 25 Hz.  The motion time is about one minute.  The trajectory is very similar to that of Fig. 2d.  The characterization of the trajectory and the comparison with experimental data deserve additional investigations since the friction influences very much the motion.  Using numerical simulations, it will be also possible to determine the over- or under-diffusive behaviour of the trimer.  

\begin{figure} [h]
\begin{center}
\includegraphics[width=10cm]{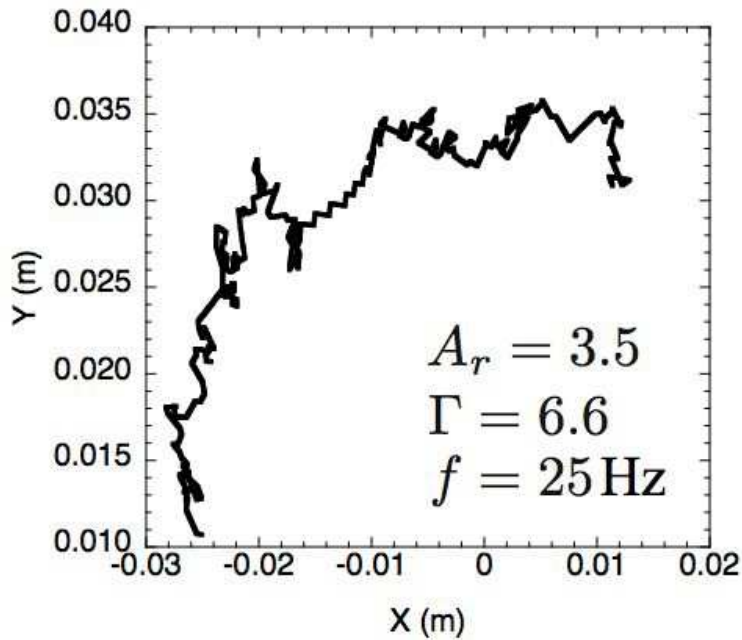}
\caption{Trajectory of the center of mass of a trimer which aspect ratio $A_r=3.5$.  The trimer is shaken at a reduced acceleration $\Gamma=6.6$ and the frequency is 25 Hz.  The duration of the simulation is about a minute.}
\end{center}

\end{figure}

\section{Conclusions}
The complexity of the object that has been placed on a shaker has been increased.  Starting from the bouncing ball, to the dimer and finishing by the trimer, the number of contacts between the object and the vibrating plate is increased.  We found that several periodical modes can be observed for the trimer like rotation, period-2 and period-3 modes.  The modes appear at a given reduced acceleration whatever the forcing frequency.   High-speed camera recordings have allowed to observe that the contact between the trimer and the plate is persistent between two successive jumps when the trimer bounces along the period-2 or period-3 modes.   This is due to the fact that at the moment of the impact the relative speed between the trimer and the plate is zero.  It results that the memory of the jump previous to the contact is completely erased.  The conditions for such a behaviour has been modeled and we have shown that period-2 and period-3 are obtained for determined reduced accelerations $\Gamma_{n}$ given by $\sqrt{(2 n+1)^2 \pi^2 /4+1}$ for $n=1$ and $n=2$ respectively.  It is remarkable that these conditions do not consider the coefficient of restitution, the friction and the shape of the particle.

The rotation modes have been characterized.  The period of rotation is proportional to the forcing frequency in the rotation-1 mode.  This result is consistent with dimer measurements and theory\cite{PRL}.  These modes occur in stability windows of the bifurcation diagram that can be reproduced numerically.  The main factor of dissipation has been found to be the coefficient of friction of the trimer with the plate.  

Finally, a self-propelled particle behaviour is found for the trimer when the condition of chaos is encountered.  Numerical simulations have shown that the stability windows and bifurcation are very robust even when the coefficient of restitution and the friction are changed.

SD would like to thank FNRS for financial support. T. Gilet is acknowledged for fruitful discussions. Part of this work has been financed by "The Interuniversity Attraction Pole INANOMAT" (IAP P6/17).

\end{document}